\documentclass[12pt]{book}

\usepackage[dvips]{graphicx,color}
\usepackage{makeidx,tsukuba}
\usepackage{rotating}

\makeauthorindex
\makeindex

\begin{document}

\BookTitle{\itshape The 28th International Cosmic Ray Conference}
\CopyRight{\copyright 2003 by Universal Academy Press, Inc.}
\pagenumbering{arabic}

\chapter{Search for Very High Energy Gamma Rays from an X-ray Selected Blazar
Sample}

\author{%
%
%
I.~de~la~Calle~Perez,$^{1,2}$ I.H.~Bond, P.J.~Boyle, S.M.~Bradbury,
J.H.~Buckley, D.~Carter-Lewis, O.~Celik, W.~Cui, M.~Daniel, M.~D'Vali,
C.~Duke, A.~Falcone, D.J.~Fegan, S.J.~Fegan, J.P.~Finley, L.F.~Fortson,
J.~Gaidos, S.~Gammell, K.~Gibbs, G.H.~Gillanders, J.~Grube, J.~Hall,
T.A.~Hall, D.~Hanna, A.M.~Hillas, J.~Holder, D.~Horan, A.~Jarvis, M.~Jordan,
G.E.~Kenny, M.~Kertzman, D.~Kieda, J.~Kildea, J.~Knapp, K.~Kosack,
H.~Krawczynski, F.~Krennrich, M.J.~Lang, S.~LeBohec, E.~Linton,
J.~Lloyd-Evans, A.~Milovanovic, P.~Moriarty, D.~Muller, T.~Nagai, S.~Nolan,
R.A.~Ong, R.~Pallassini, D.~Petry, B.~Power-Mooney, J.~Quinn, M.~Quinn,
K.~Ragan, P.~Rebillot, P.T.~Reynolds, H.J.~Rose, M.~Schroedter, G.~Sembroski,
S.P.~Swordy, A.~Syson, V.V.~Vassiliev, S.P.~Wakely, G.~Walker, T.C.~Weekes,
J.~Zweerink \\ {\it (1) Department of Physics, University of Leeds, Leeds, LS2
9JT, Yorkshire, England, UK}\\ {\it (2) The VERITAS Collaboration--see
S.P.Wakely's paper} ``The VERITAS Prototype'' {\it from these proceedings for
affiliations}}

\section*{Abstract}
In recent years, blazar surveys at radio and X-ray energies have greatly
increased our understanding of this type of active galaxy. The combination of
multi-wavelength data has shown that blazars follow a well defined sequence in
terms of their broad band spectral properties. Together with increasingly
detailed emission models, this information has provided not only tools with
which to identify potential sources of TeV emission but also predictions of
their gamma-ray flux. A list of such candidates has been used in this work
to investigate the best targets for TeV observations. Observations reported
here have resulted in upper limits which do not conflict with the latest model
predictions.

\section{Introduction}

BL Lac objects, a blazar subclass, are the preferred targets of ground-based
observations with atmospheric \v{C}erenkov telescopes since they have the
second peak of their spectral energy distribution (SED) extending well into
the TeV domain. The limited field of view of \v{C}erenkov telescopes and their
low duty cycles require a priori selection of target objects. Here our
candidates have been selected following the work of Costamante $\&$ Ghisellini
[2]. Their BL Lac catalogue is the first to provide estimates of TeV fluxes
based on model predictions. It consists of objects bright in {\it both} the
X-ray and radio bands and includes the TeV sources already detected. The
gamma-ray flux at TeV energies has been estimated by applying a
homogeneous one-zone SSC model [3], and by using the phenomenological
parameterisation of the blazar SED developed by Fossati [1] and
modified by Costamante [2] to better describe the SEDs of low power blazars.

\begin{table}[h]
 \caption{Results for the 8 objects considered. {\it z} is the
 source redshift. (a) Flux upper limit in Crab units above 390~GeV at a $97\%$
 c.l. (b) Estimated absorption of the gamma-ray photon flux between 390~GeV
 and 10~TeV (using the optical depth given in [12] and assuming a Crab-like
 source spectrum). (c) The predicted fluxes above 300~GeV as described in [2]
 (Fossati/SSC) converted to Crab units. (d) Observing time needed for a
 5$\sigma$ detection given a telescope sensitivity of
 $5.74/\sqrt{t(h)}$.\label{results_tab}}
\begin{center}
{\small
\begin{tabular}{lc|ccccc}
\hline
Source & {\it z} & Obs. Time & 
$U.L.^{(a)}$ & IR$^{(b)}$ & Flux Pred.$^{(c)}$ & Req. Time$^{(d)}$ \\

      &    &   (h)      & 
       (c.u.)                 & ($\%$)    & (c.u.)     &   (h)   \\
 
\hline
{\bf 1ES0033+595}  & 0.086 & 12.02 & 
0.11 &  57 & 0.17/0.021 &  23/1482    \\
			                          	                          	            
{\bf 1ES0120+340}  & 0.272 &  5.05 &
0.12 &  94 & 0.02/0.025 &  1135/1047  \\
			                          	                          	            
{\bf RGBJ0214+517} & 0.049 &  6.05 &  
0.17 &  38 & 0.48/0.006 &  2.8/18107  \\
			                          	                          	            
{\bf 1ES0229+200}  & 0.139 & 14.69 & 
0.11 &  74 & 0.08/0.026 &  101/968    \\
			                          	                          	            
{\bf 1ES0806+524}  & 0.138 & 18.70 & 
0.08 &  74 & 0.11/$---$ &  50/---     \\ 
			                          	                         	            
{\bf RGBJ1117+202} & 0.139 & 3.26 & 
0.21 &  74 & 0.09/0.008 &  68/10189   \\ 
			                          	                           	            
{\bf 1ES1553+113}  & 0.360 & 2.82 & 
0.19 &  98 & 0.02/0.035 &  2260/535   \\
			                          	                           	            
{\bf RGBJ1725+118} & 0.018 & 2.33 &
0.23 &  17 & 1.04/0.001 &  0.67/651240\\
							   
\hline
\end{tabular}
}
\end{center}
\end{table}

\section{Observations and Analysis}

Eight objects from this BL Lac sample were observed with the Whipple 10-m
gamma-ray telescope [5] during 2001-2002. The selection was based mainly on
the source redshift ({\it z} $\leq 0.2$) and requiring the predicted flux above
300~GeV, according to [1], to be $\geq 10\%$ of the Crab Nebula flux to keep
observing times below $\sim$~50~hours. 1ES0120+340 and 1ES1553+113, although
at a larger redshift and with flux predictions at $\sim~2\%$ the Crab Nebula
flux, were also included on the basis of their extreme nature [6], both having
many similar broadband properties to 1ES1426+428. Table \ref{results_tab}
lists the predicted flux values as taken from [2]. Data for which only ON
source observations are available have been analysed using a background
estimate obtained from OFF source data from different regions of the sky,
which are 'matched' to the ON source data [8]. This analysis includes software
padding. A set of image parameter cuts [5], optimized on Crab Nebula data,
has been applied to select gamma-ray candidate events [7]. A combination of
genuine and matched ON/OFF pairs are used to establish the statistical
significance of any excess as described in [9].

\section{Results}

\begin{sidewaysfigure}[t]
  \begin{center}
    \includegraphics[height=22.0pc]{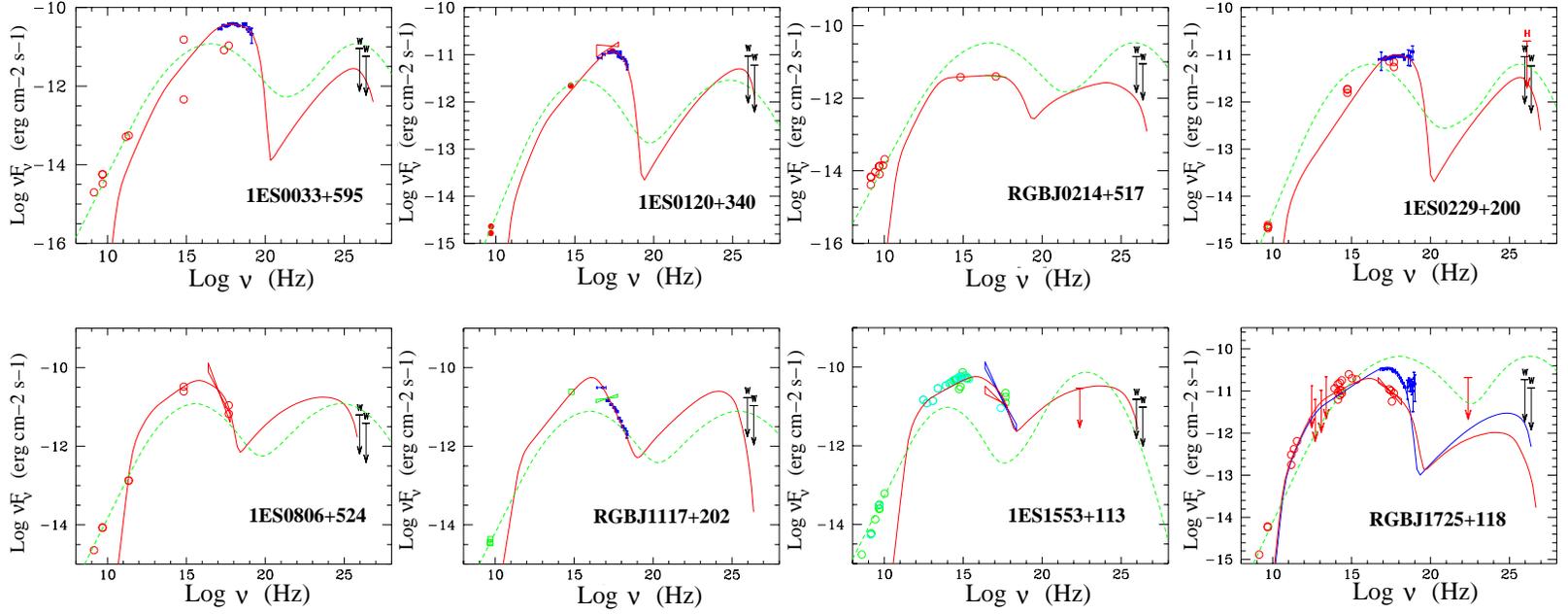}
  \end{center}
  \vspace{-0.5pc}
  \caption{SSC model (solid line)
and phenomenological parameterisation of Fossati [1] as modified by Costamante
[2] (dashed line). Flux upper limits obtained in this work are represented by
arrows labeled as {\it W} at two energies, 390~GeV and 1~TeV. In the
1ES0229+200 SED plot {\it H} stands for HEGRA. See [2] for more details 
(Figures courtesy of L. Costamante).  \label{SSC_models}}
\end{sidewaysfigure}

No evidence for TeV emission was found from any of the eight objects over
short or long time scales. Source flux upper limits were derived
using the Helene method [10] and referred to contemporaneous Crab Nebula
observations as described in [11]. Our upper limits in terms of energy
density $\nu~F_\nu$ are presented in Fig.\ref{SSC_models} These upper
limits apply only to the epoch of the reported observations.

\section{Summary}

Our flux upper limits are consistent with current SSC models and for 4 of the
objects are below the flux predicted by the Fossati approach. Due to the high
variability that this type of object presents, our non-detections should not
discourage future observations of objects selected from this sample, as the
detection of 1ES1959+650 at TeV energies indicates [4]. This object was
predicted as a TeV source, and although its predicted flux $>$0.3~TeV was only
3~mCrab according to the SSC model (630~mCrab using the Fossati approach), it
was detected at TeV energies at a level of up to 5~Crab. With the VERITAS-4
telescope array, with a sensitivity to a Crab-like source of 5~mCrab in
50~hours (5$\sigma$ detection), the model predictions will be strongly tested
in just a few hours.

\section{Acknowledgments}

We acknowledge the technical assistance of E. Roache and J. Melnick. This
research is supported by grants from the U. S. Department of Energy, by
Enterprise Ireland and by PPARC in the UK.

\section{References}
\re
1.\ Fossati G.\ et al.\ 1998, MNRAS 299, 433
\re
2.\ Costamante L. $\&$ Ghisellini G.\ 2002, A$\&$A 384, 56
\re
3.\ Ghisellini G., Celotti A. $\&$ Costamante L.\ 2002, A$\&$A 386, 833
\re
4.\ Holder J.\ et al.\ 2002, ApJL 583, L9
\re
5.\ Finley P. J.\ et al.\ 2001, in Proc. 27${th}$ ICRC 7, 2827, Hamburg, Germany
\re
6.\ Ghisellini G.\ 1999, Astroparticle Physics 11, 11
\re
7.\ Reynolds P. T.\ et al.\ 1993, ApJ 404, 206
\re
8.\ de la Calle Perez I.\ et al.\ {\it in preparation}
\re
9.\ Catanese M. et al.\ 1998, ApJ 501, 616
\re
10.\ Helene O.\ 1983, NIM 212, 319
\re
11.\ Aharonian F. A.\ et al.\ 2000, A$\&$A 353, 847
\re
12.\ de Jager O. C. $\&$ Stecker F. W.\ 2002, ApJ 566, 738

\endofpaper
\end{document}